\documentclass{ws-ijmpe}
\usepackage[super,compress]{cite}

\begin{document}

\markboth{M. Thoennessen}{2022 Update of the Discoveries of Isotopes}

\catchline{}{}{}{}{}

\title{2022 UPDATE OF THE DISCOVERIES OF NUCLIDES}

\author{\footnotesize M. THOENNESSEN}

\address{Facility for Rare Isotope Beams and \\
Department of Physics \& Astronomy \\
Michigan State University\\
East Lansing, Michigan 48824, USA\\
thoennessen@nscl.msu.edu}

\maketitle

\begin{history}
\received{Day Month Year}
\revised{Day Month Year}
\end{history}

\begin{abstract}
The 2022 update of the discovery of nuclide project is presented. It is the first update in four years, and 36 new nuclides were observed for the first time during 2019-2022. Isotopes that have so far only been published in conference proceedings or internal reports are also listed. 
\end{abstract}


\ccode{PACS numbers: 21.10.-k, 29.87.+g}


\section{Introduction}

This is the seventh update of the isotope discovery project which was originally published in a series of papers in Atomic Data and Nuclear Data Tables from 2009 through 2013 (see for example the first\cite{2009Gin01} and last\cite{2013Fry01} papers). Two summary papers were published in 2012 and 2013 in Nuclear Physics News\cite{2012Tho03} and Reports on Progress in Physics,\cite{2013Tho02} respectively, followed by annual updates in 2014,\cite{2014Tho01} 2015,\cite{2015Tho01} 2016,\cite{2016Tho02} 2017,\cite{2017Tho01} 2018,\cite{2018Tho01} and 2019.\cite{2019Tho01} No updates were published since then because of the small number of isotopes discovered during these years. In 2016 a description of the discoveries from a historical perspective was published in the book ``The Discovery of Isotopes -- A complete Compilation''.\cite{2016Tho01}

\section{New discoveries in 2019}
\label{New2019}

In 2019, the discoveries of six new nuclides were reported in refereed journals. At that time this was the smallest number of isotopes discovered in a year (tying 2014) since 1944 when only two new isotopes were reported. These six new isotopes are located on the proton-rich side of the chart of nuclides. Three of them were produced with secondary beams following projectile fragmentation, while the other three were populated using fusion evaporation reactions.  Table \ref{2019Isotopes} lists details of the discoveries including the production method. 

\begin{table}[pt]
\tbl{New nuclides reported in 2019. The nuclides are listed with the first author, submission date, and reference of the publication, the laboratory where the experiment was performed, and the production method (FE = fusion evaporation, SB = secondary beams). \label{2019Isotopes}}
{\begin{tabular}{@{}llrclc@{}} \toprule 
Nuclide(s) & First Author & Subm. Date & Ref. & Laboratory & Type \\ \colrule
$^{11}$O & T. B. Webb & 12/20/2018 & \refcite{2019Web01} & MSU & SB  \\
$^{220}$Np & Z. Y. Zhang & 1/9/2019 & \refcite{2019Zha01} & Lanzhou & FE  \\ 
$^{165}$Pt, $^{170}$Hg & J. Hilton & 5/13/2019 & \refcite{2019Hil01} & Jyv\"askyl\"a & FE \\
$^{68}$Br & K. Wimmer & 5/15/2019 & \refcite{2019Wim01} & RIKEN &SB  \\
$^{31}$K & D. Kostyleva & 5/22/2019 & \refcite{2019Kos01} & GSI & SB \\
\botrule
\end{tabular}}
\end{table}

Webb et al. reported the discovery of $^{11}$O in the paper entitled ``First Observation of Unbound $^{11}$O, the Mirror of the Halo Nucleus $^{11}$Li''.\cite{2019Web01} A secondary $^{13}$O beam was produced from a 150 MeV/nucleon $^{16}$O beam at the National Superconducting Cyclotron Facility at Michigan State University. The $^{11}$O isotopes were then populated via two-neutron knockout reaction on a 1-mm thick $^9$Be target and identified by measuring charged reaction products in the High Resolution Array (HiRA) consisting of 14 $\Delta$E-E (Si-CsI(Tl)) detectors. ``The energies $Q_{2p}$ (and widths) of the four lowest-lying resonant states in $^{11}$O obtained with $V_0$ optimized to the observed energy spectrum are 4.16(1.30) MeV for 3/2$^-_1$, 4.65 (1.06) MeV for 5/2$^+_1$, 4.85(1.33) MeV for 3/2$^-_2$, and 6.28(1.96) MeV for 5/2$^+_2$.''

$^{220}$Np was discovered by Zhang et al. in ``New Isotope $^{220}$Np: Probing the Robustness of the N = 126 Shell Closure in Neptunium''.\cite{2019Zha01} The Sector-Focusing Cyclotron (SFC) at the Heavy Ion Research Facility in Lanzhou (HIRFL) was used to accelerate a $^{40}$Ar beam to 201 MeV and $^{220}$Np was populated in the fusion evaporation reaction $^{185}$Re($^{40}$Ar,5n). Recoiled evaporation residues were separated with the gas-filled separator SHANS and implanted into three position-sensitive 16-strip detectors. $^{220}$Np was identified by correlating implantation events with subsequent detection of $\alpha$ particles. ``Based on the measurement of the correlated $\alpha$-decay chains, the decay properties of $^{220}$Np with E$_\alpha$ = 10040(18) keV and T$_{1/2}$ = 25$^{+14}_{-7} \mu s$ were determined, which are in good agreement with theoretical predictions.''

Hilton et al. discovered $^{165}$Pt and $^{170}$Hg in ``$\alpha$-spectroscopy studies of the new nuclides $^{165}$Pt and $^{170}$Hg''.\cite{2019Hil01} $^{78}$Kr beams from the University of Jyv\"askyl\"a K130 cyclotron irradiated $^{92}$Mo (at 418 MeV) and $^{96}$Ru (at 390 MeV) targets to produce $^{165}$Pt and $^{170}$Hg in 5n and 4n fusion evaporation reactions, respectively. The mass analyzing recoil apparatus vacuum (MARA) was used to separate the residues which were then implanted in a double-sided silicon strip detector. The isotopes were identified from correlations with the subsequent radioactive decays. ``For $^{170}$Hg an $\alpha$-particle energy of E$_\alpha$ = 7590(30) keV and half-life of t$_{1/2}$ = 0.08$^{+0.40}_{-0.04} ms$ were deduced, while for $^{165}$Pt the corresponding values were 7272(14) keV and 0.26$^{+0.26}_{-0.09} ms$.'' 

In ``Discovery of $^{68}$Br in secondary reactions of radioactive beams'', Wimmer et al. described the first observation of $^{68}$Br.\cite{2019Wim01} A 345 MeV/nucleon $^{78}$Kr beam from the Radioactive Isotope Beam Factory at RIKEN was used to produce secondary beams of $^{70}$Br, $^{71}$Kr, and $^{72}$Kr with the BigRIPS separator. These beams impinged on a secondary 703(7) mg/cm$^2$ Be target at about 170 MeV/nucleon and the reaction products were separated and analyzed in the ZeroDegree spectrometer by measuring their energy loss, time-of-flight and magnetic rigidity. ``...12 events are observed at Z=35, A/q =1.943, corresponding to $^{68}$Br. After correction for the detection efficiency and the transmission through the spectrometers, the yield of $^{68}$Br amounts to 14.7(50)(18) with statistical and systematic uncertainties, respectively.'' A previous search using projectile fragmentation reactions did not observe any $^{68}$Br events.\cite{2014San01}

The discovery of $^{31}$K was reported by Kostyleva et al. in ``Towards the Limits of Existence of Nuclear Structure: Observation and First Spectroscopy of the Isotope $^{31}$K by Measuring Its Three-Proton Decay''.\cite{2019Kos01} A secondary beam of 620 MeV/nucleon $^{31}$Ar was produced from a primary 885 MeV/nucleon $^{36}$Ar beam at the SIS-FRS facility at GSI and impinged on a 27-mm thick $^9$Be target. $^{31}$K was populated in charge-exchange reactions and was identified in the second half of the FRS by reconstructing the reaction products: $^{28}$S and three protons. ``The energies of the previously unknown ground and excited states of $^{31}$K have been determined. This provides its 3p separation energy value $S_{3p}$ of $-4.6(2)$ MeV.''

\section{New discoveries in 2020}
\label{New2020}

In 2020, even fewer isotopes were discovered. The four newly reported isotopes were the lowest since 1944. Only four times since Soddy received the Nobel prize for the discovery of isotopes in 1922 were fewer isotopes detected in any given year: 1926 (2), 1928 (1), 1942 (3) and 1944 (2). All four isotopes were populated using fusion evaporation reactions and three are isotopes of transuranium elements. This was the first year since 2006 that no isotopes were discovered in projectile fragmentation reactions. It is worth noting that the four isotopes were discovered in four different countries. Table \ref{2020Isotopes} lists details of the discoveries including the production method. 

$^{244}$Md was first reported in ``Identification of the New Isotope $^{244}$Md'' by Pore et al.\cite{2020Por01} The isotope was produced in the $^{209}$Bi($^{40}$Ar,5n) reaction with a 220 MeV $^{40}$Ar beam accelerated in the 88-inch cyclotron at Lawrence Berkeley National Laboratory. The $^{209}$Bi target had an average thickness of 0.500 mg/cm$^2$. Residues were separated and identified with the Berkeley Gas-filled separator and FIONA (For the Identification Of Nuclide A) where their decays were detected after implementation into 32$\times$32 strip Double-sided Silicon Strip Detectors (DSSD). ``The isotope $^{244}$Md is reported to have one, possibly two, $\alpha$-decaying states with $\alpha$ energies of 8.66(2) and 8.31(2) MeV and half-lives of 0.4$^{+0.4}_{-0.1}$ and $\sim$6 s, respectively.''  Less than four months later, Khuyagbaatar et al. independently submitted their results reporting the observation of $^{244}$Md with $\alpha$-particle energies of 8.73–8.86 MeV and a half-life of 0.30$^{+0.19}_{-0.09}$ s.\cite{2020Khu02} In a subsequent paper He\ss berger et al. claim that the decays presented by Pore et al. should be reassigned to $^{245}$Md (with one event each to  $^{245}$Fm and $^{246}$Md).\cite{2021Hes01} This discrepancy has not been resolved.

In the paper ``$\alpha$ decay of $^{243}$Fm$_{143}$ and $^{245}$Fm$_{145}$, and of their daughter nuclei'' Khuyagbaatar et al. described the first observation of $^{235}$Cm.\cite{2020Khu01} The linear accelerator UNILAC was used to accelerate $^{40}$Ar to 185$-$204 MeV and impinge on a $\sim$400 $\mu$g/cm$^2$ lead sulfide target. Evaporation residues were separated with the velocity filter SHIP and implanted in a position sensitive 16-strop silicon detector. $^{243}$Fm was formed in the reaction $^{40}$Ar + $^{206}$Pb: ``A signature for detection of the hitherto unknown $^{235}$Cm was found in the $\alpha$-decay chains from $^{243}$Fm. Two groups of $\alpha$ events with average energies of 6.69(2) MeV and 7.01(2) MeV and with a half-life of T$_{1/2}$ = 300$^{+250}_{-100}$ s are suggested to originate from $^{235}$Cm.'' 

Ma et al. discovered $^{222}$Np in ``Short-Lived $\alpha$-Emitting Isotope $^{222}$Np and the Stability of the N = 126 Magic Shell''.\cite{2020Ma01} A 250 $\mu$g/cm$^2$ thick $^{187}$Re target was bombarded with a 198.7 MeV $^{40}$Ar beam at the Heavy Ion Research Facility in Lanzhou, China. $^{222}$Np was formed in the reaction $^{187}$Re($^{40}$Ar,5n). The gas-filled separator SHANS was used to separate the reaction products. The isotopes of interest were implanted in three position-sensitive 16-strip detectors which also recorded subsequent $\alpha$ decays. ``The decay properties of $^{222}$Np with E$_\alpha$ = 10016(33) keV and T$_{1/2}$ = 380$^{+260}_{-110}$ ns were determined experimentally.'' 

\begin{table}[pt]
\tbl{New nuclides reported in 2020. The nuclides are listed with the first author, submission date, and reference of the publication, the laboratory where the experiment was performed, and the production method (FE = fusion evaporation). \label{2020Isotopes}}
{\begin{tabular}{@{}llrclc@{}} \toprule 
Nuclide(s) & First Author & Subm. Date & Ref. & Laboratory & Type \\ \colrule
$^{244}$Md & J. L. Pore & 1/31/2020 & \refcite{2020Por01} & Berkeley & FE \\
$^{235}$Cm & J. Khuyagbaatar &2/26/2020 & \refcite{2020Khu01} & GSI & FE \\ 
$^{222}$Np & L. Ma & 5/8/2020 & \refcite{2020Ma01} & Lanzhou & FE \\ 
$^{211}$Pa & K. Auranen & 6/23/2020 & \refcite{2020Aur01} & Jyv\"askyl\"a & FE \\ 
\botrule
\end{tabular}}
\end{table}

``Exploring the boundaries of the nuclear landscape: $\alpha$-decay properties of $^{211}$Pa''.\cite{2020Aur01} The University of Jyv\"askyl\"a K-130 cyclotron accelerated an $^{36}$Ar beam to 178$-$214 MeV and irradiated 1000 and 450 $\mu$g/cm$^2$ thick tantalum targets. Reaction products were separated with the recoil ion transport unit (RITU) gas-filled separator and stopped in the implantation detector of the GREAT spectrometer which also recorded subsequent $\alpha$ decays. ``Using a fusion-evaporation reaction and a gas-filled recoil separator we have produced and, for the first time, identified rigorously the new isotope $^{211}$Pa. It was found to $\alpha$ decay with a half-life and $\alpha$-particle energy of 3.8$^{+4.6}_{-1.4}$ ms and 8320(40) keV, respectively, suggesting a favored $\alpha$ decay.''

\section{New discoveries in 2021}
\label{New2021}

In 2021, the discoveries of 15 new nuclides were reported in refereed journals. Table \ref{2021Isotopes} lists details of the discoveries including the production method. 

The neutron-rich nuclei $^{101}$Br, $^{102}$Kr, $^{105}$Rb, $^{106}$Rb, $^{108}$Sr,  $^{110}$Y, $^{111}$Y, $^{114}$Zr, and $^{117}$Nb were discovered by Sumikama et al. in the paper entitled ``Observation of new neutron-rich isotopes in the vicinity of $^{110}$Zr''.\cite{2021Sum01} The isotopes were produced by in-flight fission from a 345 MeV/nucleon $^{238}$U at the RIKEN Radioactive Isotope Beam Factory (RIBF) and separated and identified with the large-acceptance two-stage fragment separator BigRIPS and the ZeroDegree spectrometer. ``Ten candidates for previously unreported neutron-rich isotopes were produced, namely, events corresponding to fully stripped ions of $^{99,101}$Br, $^{102}$Kr, $^{105,106}$Rb, $^{108}$Sr, $^{110,111}$Y, $^{114}$Zr, and $^{117}$Nb. The A/Q values of new-isotope events were consistent with those extrapolated from other isotopes.'' Of these isotopes $^{99}$Br was not claimed as a new isotope as it was not possible to determine its production cross section.

\begin{table}[b]
\tbl{New nuclides reported in 2021. The nuclides are listed with the first author, submission date, and reference of the publication, the laboratory where the experiment was performed, and the production method (PF = projectile fragmentation, FE = fusion evaporation, SB = secondary beams, SP = spallation). \label{2021Isotopes}}
{\begin{tabular}{@{}llrclc@{}} \toprule 
Nuclide(s) & First Author & Subm. Date & Ref. & Laboratory & Type \\ \colrule
$^{101}$Br, $^{102}$Kr, $^{105}$Rb, & T. Sumikama & 7/6/2020 & \refcite{2021Sum01} & RIKEN & PF \\ 
$^{106}$Rb, $^{108}$Sr, $^{110}$Y, &  &  & &  \\
$^{111}$Y, $^{114}$Zr, $^{117}$Nb &  &  & &  \\ 
 $^{280}$Ds & A. Samark-Roth & 11/16/2020 & \refcite{2021Sam01} & GSI & FE \\
 $^{13}$F & R. J. Charity & 11/16/2020 & \refcite{2021Cha01} & MSU & SB \\
 $^{249}$No & J. Khuyagbaatar & 1/3/2021  & \refcite{2021Khu01} & GSI & FE \\
 $^{214}$U & Z. Y. Zhang & 1/15/2021 & \refcite{2021Zha01} & Lanzhou & FE \\
 $^{150}$Yb & S. Beck & 4/13/2021 & \refcite{2021Bec01} & TRIUMF & SP \\
  $^{18}$Mg & Y. Jin & 8/30/2021 & \refcite{2021Jin01} & MSU & SB \\
\botrule
\end{tabular}}
\end{table}

The first observation of $^{280}$Ds was reported by S\aa mark-Roth et al. in ``Spectroscopy Along Flerovium Decay Chains: Discovery of $^{280}$Ds and an Excited State in $^{282}$Cn''.\cite{2021Sam01} A $^{48}$Ca beam accelerated to 6.021(2) MeV/nucleon by the Universal Linear Accelerator (UNILAC) at the GSI Helmholtzzentrum f\"ur Schwerionenforschung, in Darmstadt, Germany, was delivered to a rotating target wheel of enriched $^{244}$Pu deposited on titanium foils. Reaction products were separated by the recoil separator TASCA and identified in the TASISpec decay station. ``In one case, a Q$_\alpha$ = 9.46(1)-MeV decay from $^{284}$Cn into $^{280}$Ds was observed, with $^{280}$Ds fissioning after only 518 $\mu$s.'' The spontaneous fission of $^{280}$Ds had previously been reported as tentative\cite{2015Mor01} or uncertain.\cite{2017Kaj01}

Charity et al. described the discovery of $^{13}$F in ``Observation of the Exotic Isotope $^{13}$F Located Four Neutrons beyond the Proton Drip Line''.\cite{2021Cha01} The Coupled Cyclotron Facility at the National Superconducting Cyclotron Laboratory at Michigan State University produced a 69.5 MeV/nucleon secondary $^{13}$O beam from a primary 150 MeV/nucleon $^{16}$O beam. The $^{13}$O beam impinged on a 1-mm thick beryllium target after being purified with the A1900 and the radio frequency fragment separators. Charged fragments from the reactions were detected in the high resolution array (HiRA) and the decay energy of $^{13}$F was measured from the 3p + $^{10}$C events: ``The invariant-mass distribution of detected 3p + $^{10}$C events displays a 1.01(27)-MeV-wide peak at a decay energy of 7.06(9) MeV sitting on a broad background. Based on predictions of Fortune and Sherr [...], this peak corresponds to the first 5/2$^+$ excited state in $^{13}$F.''

$^{249}$No was discovered by Khuyagbaatar et al. in ``Spontaneous fission instability of the neutron-deficient No and Rf isotopes: The new isotope $^{249}$No''.\cite{2021Khu01} A 234.3 MeV $^{50}$Ti beam from the Universal Linear Accelerator UNILAC at GSI, Darmstadt, Germany, impinged on a 0.6 mg/cm$^2$ thick enriched $^{204}$PbS target. Fusion-evaporation residues and their correlated subsequent $\alpha$ decays were analyzed with the gas-filled transActinide Separator and Chemistry Apparatus (TASCA). ``For the fission and $\alpha$-decaying state in $^{253}$Rf a partial $\alpha$-decay half-life (T$_\alpha$) of $\sim$100 ms was deduced with its $\sim$12.5\% $\alpha$ branching. The state populated in the new isotope $^{249}$No decays by $\alpha$-particle emission with an energy of 9.06(3) MeV and a half-life of 15$^{+74}_{-7}$ ms.'' Less than 2-months later Svirikhin et al. independently submitted their result of $^{249}$No reporting an $\alpha$ decay energy of 9129 keV and the half-life of 38.1 $\pm$ 2.5 ms.\cite{2021Svi01}

In the paper ``New $\alpha$-Emitting Isotope $^{214}$U and Abnormal Enhancement of $\alpha$-Particle Clustering in Lightest Uranium Isotopes'' Zhang et al. reported the first observation of $^{214}$U.\cite{2021Zha01} The Heavy Ion Research Facility in Lanzhou (HIRFL), China was used to accelerate a $^{36}$Ar beam of 184 MeV which then was delivered on to 300$-$350 $\mu$g/cm$^2$ $^{182}$W sputtered on carbon foils. Evaporation residues were separated with the gas-filled recoil separator Spectrometer for Heavy Atoms and Nuclear Structure (SHANS) and implanted in three 16-strip position-sensitive silicon detectors which also recorded correlated subsequent $\alpha$ decays. ``Two decay events in Fig. [...] were assigned to the new isotope $^{214}$U unambiguously. [...] Based on these measurements, the mean $\alpha$-particle energy and half-life of $^{214}$U were determined to be 8533(18) keV and 0.52$^{+0.95}_{-0.21}$ ms, respectively,...''

Beck et al. observed $^{150}$Yb for the first time as described in ``Mass Measurements of Neutron-Deficient Yb Isotopes and Nuclear Structure at the Extreme Proton-Rich Side of the N = 82 Shell''.\cite{2021Bec01} A 480 MeV proton beam from the TRIUMF cyclotron in Vancouver, Canada, irradiated a tantalum target. Spallation reaction products were extracted and separated with the Isotope Separator and Accelerator (ISCA) facility and transported to TRIUMF’s Ion Trap for Atomic and Nuclear Science (TITAN) facility. Masses of the isotopes of interest were measured with the multiple-reflection time-of-flight mass spectrometer (MR-TOF-MS). "The masses of $^{150}$Yb and $^{153}$Yb and the excitation energy of $^{151}$Yb$^m$ were measured for the first time.'' The authors did not consider their results as the discovery of $^{150}$Y because the observation of this isotope in a fragmentation reaction had been reported previously in a conference proceeding.\cite{2000Sou01} However, it was not subsequently published in a refereed journal.

The discovery of $^{18}$Mg was reported by Jin et al. in the paper ``First Observation of the Four-Proton Unbound Nucleus $^{18}$Mg''.\cite{2021Jin01} A secondary 103 MeV/nucleon $^{20}$Mg beam was produced from a primary 170 MeV/nucleon $^{24}$Mg beam at the Coupled Cyclotron Facility of the National Superconducting Cyclotron Laboratory at Michigan State University. After separation with the A1900 fragment separator the $^{20}$Mg impinged on a 1-mm thick $^9$Be target and produced $^{18}$Mg in two-neutron knockout reactions which then decayed into $^{14}$O and four protons. The $^{14}$O was detected around zero degrees in an orthogonal array of scintillating fiber ribbons. Protons were recorded in an annular double-sided silicon-strip detector backed by an annular array of CsI(Tl) crystals. ``We have observed, for the first time, $^{18}$Mg via its decay into 4p + $^{14}$O. The ground-state decay energy was found to be E$_T$ = 4.865(34) MeV.''

\section{New discoveries in 2022}
\label{New2022}
Eleven isotopes were discovered  and reported in refereed journals in 2022. Table \ref{2022Isotopes} lists details of the discoveries including the production method. 

Auranen et al. announced the discovery of $^{149}$Lu in ``Nanosecond-Scale Proton Emission from Strongly Oblate-Deformed $^{149}$Lu''.\cite{2022Aur01} The University of Jyv\"askyl\"a K130 cyclotron provided $^{58}$Ni beams of 310 and 320 MeV which were focused on 170 $\mu$g/cm$^2$ thick enriched $^{96}$Ru evaporated on an aluminum foil. Evaporation residues were separated with the Mass Analyzing Recoil Apparatus (MARA) and deposited in a double-sided silicon strip detector (DSSD) located behind a multiwire proportional counter. Emitted protons were measured by recording a long wave form sample of each event from one side of the DSSD. ``Using the fusion-evaporation reaction $^{96}$Ru($^{58}$Ni,p4n)$^{149}$Lu and the MARA vacuum-mode recoil separator, a new proton-emitting isotope $^{149}$Lu has been identified. The measured decay Q value of 1920(20) keV is the highest measured for a ground-state proton decay, and it naturally leads to the shortest directly measured half-life of 450$^{+170}_{-100}$ ns for a ground-state proton emitter.''

$^{207}$Th was discovered in the paper ``New isotope $^{207}$Th and odd-even staggering in alpha-decay energies for nuclei with Z $>$ 82 and N $<$ 126'' by Yang et al.\cite{2022Yan01}  Isotopically enriched $^{176}$Hf (116$-$360$\mu$g/cm$^2$) were irradiated with 197$-$199 MeV $^{36}$Ar beams accelerated by the Sector Focusing Cyclotron of the Heavy Ion Research Facility in Lanzhou (HIRFL), China. Reaction products were separated by the Spectrometer for Heavy Atoms and Nuclear Structure (SHANS) and implanted in three 300-$\mu$m-thick position-sensitive silicon strip detectors (PSSDs). Subsequent correlated $\alpha$ decays were recorded in these and eight additional non-position-sensitive silicon detectors surrounding the implantation detectors. ``The $\alpha$ decay of $^{207}$Th, measured with an $\alpha$-particle energy of 8167(21) keV and a half-life of 9.7$^{+46.6}_{-4.4}$ ms, is assigned to originate from ground state.''

The first observation of $^{264}$Lr was reported by Oganessian et al. in the paper ``First experiment at the Super Heavy Element Factory: High cross section of $^{288}$Mc in the $^{243}$Am + $^{48}$Ca reaction and identification of new isotope $^{264}$Lr''.\cite{2022Oga01} The new DC280 cyclotron at the Super Heavy Element Factory (SHE Factory) delivered $\sim$240 MeV $^{48}$Ca beams to 0.36 and 0.38 mg/cm$^2$ thick enriched $^{243}$Am targets. Fusion-evaporation residues were separated with the new DGFRS-2 separator where they were identified in two multiwire proportional chambers (MWPC) and two double-sided strip detectors (DSSD). The implantation events were correlated with subsequent $\alpha$ decays which were recorded in the DSSDs as well as eight strip detectors forming a box around the DSSD. ``The $\alpha$ decay of $^{268}$Db with an energy of 7.6$-$8.0 MeV, half-life of 16$^{+6}_{-4}$ h, and an $\alpha$ branch of 55$^{+20}_{-15}$\% was registered for the first time, and a new spontaneously fissioning isotope $^{264}$Lr with a half-life of 4.9$^{+2.1}_{-1.3}$ h was synthesized.''

\begin{table}[pt]
\tbl{New nuclides reported in 2022. The nuclides are listed with the first author, submission date, and reference of the publication, the laboratory where the experiment was performed, and the production method (PF = projectile fragmentation, FE = fusion evaporation). \label{2022Isotopes}}
{\begin{tabular}{@{}llrclc@{}} \toprule 
Nuclide(s) & First Author & Subm. Date & Ref. & Laboratory & Type \\ \colrule
 $^{149}$Lu & K. Auranen & 12/15/2021 & \refcite{2022Aur01} &  Jyv\"askyl\"a  & FE \\
 $^{207}$Th & H. B. Yang & 2/7/2022 & \refcite{2022Yan01} & Lanzhou & FE \\
 $^{264}$Lr & Yu. Ts. Oganessian & 3/11/2022  & \refcite{2022Oga01} & Dubna & FE \\
 $^{166}$Pm, $^{168}$Sm, & G. G. Kiss & 4/29/2022 & \refcite{2022Kis01} &  RIKEN & PF \\
 $^{170}$Eu, $^{172}$Gd &  &  &  &  &  \\
 $^{204}$Ac & M. H. Huang & 6/27/2022 & \refcite{2022Hua01} & Lanzhou & FE \\
 $^{251}$Lr & T. Huang & 6/30/2022 & \refcite{2022Hua02} & Argonne & FE \\
 $^{39}$Na & D. S. Ahn & 9/8/2022 & \refcite{2022Ahn01} & RIKEN & PF \\
 $^{286}$Mc  & Yu. Ts. Oganessian & 10/10/2022  & \refcite{2022Oga02} & Dubna & FE \\
\botrule
\end{tabular}}
\end{table}

$^{166}$Pm, $^{168}$Sm, $^{170}$Eu, and $^{172}$Gd were discovered by G. G. Kiss et al. as reported in ``Measuring the $\beta$-decay Properties of Neutron-rich Exotic Pm, Sm, Eu, and Gd Isotopes to Constrain the Nucleosynthesis Yields in the Rare-earth Region''.\cite{2022Kis01} A 345 MeV/nucleon $^{238}$U primary beam impinged on a 5 mm thick $^9$Be target at the RIKEN Nishina Center and the fragmentation products were separated with the large-acceptance BigRIPS separator and deposited in the AIDA implantation detector which consisted of a stack of six double-sided silicon strip detectors. ``Only a few hundred (or even fewer) ions of the most neutron-rich isotopes ($^{166}$Pm, $^{167,168}$Sm, $^{170}$Eu, and $^{170-172}$Gd) were implanted in the AIDA detector.'' These isotopes had previously been reported in an RIKEN accelerator progress report.\cite{2015Fuk01}

M. H. Huang et al. discovered $^{204}$Ac in ``$\alpha$-decay of the new isotope $^{204}$Ac''.\cite{2022Hua01} The China Accelerator Facility for superheavy Elements (CAFE2) produced a 200 MeV $^{40}$Ca beam which was focused on a rotating target wheel of twenty 450 $\mu$g/cm$^2$ $^{169}$Tm targets. Evaporation residues from the reaction $^{169}$Tm($^{40}$Ca,5n)$^{204}$Ac were detected and identified with the gas-filled recoil separator SHANS2 (Spectrometer for Heavy Atoms and Nuclear Structure-2). The deposited ions and their subsequent $\alpha$-decay were measured with silicon strip detectors. ``Nineteen ER - $\alpha_1 - \alpha_2 - \alpha_3$ decay chains which include eleven chains observed at SHANS2 were assigned to $^{204}$Ac.''

In the paper ``Discovery of the new isotope $^{251}$Lr: the impact of the hexacontetrapole deformation on single-proton orbital energies near the Z=100 deformed shell gap''  T. Huang et al. present the first observation of $^{251}$Lr.\cite{2022Hua02} The ATLAS linear accelerator at the Argonne National Laboratory delivered a 237 MeV $^{50}$Ti beam to a 0.5 mg/cm$^2$ enriched $^{203}$Tl target forming $^{251}$Lr in the 2n evaporation reaction $^{203}$Tl,($^{50}$Ti,2n). The evaporation residues were separated with the Argonne Gas-Filled Analyzer (AGFA) and implanted in a double-sided silicon stop detector which also recorded subsequent correlated $\alpha$ decays. ``Two $\alpha$-decay activities with energies of 9210(19) and 9246(19) keV and the half-lives of 42$^{+42}_{-14}$ ms and 24.4$^{+7.0}_{-4.5}$ ms were observed which were followed by the known $\alpha$ decays of $^{247}$Md and $^{243}$Es. They are interpreted as originating from the 1/2$^-$[521] and 7/2$^-$[514] single-proton Nilsson states in the hitherto unknown isotope $^{251}$Lr.''

The discovery of $^{39}$Na was discussed in ``Discovery of $^{39}$Na'' by Ahn et al.\cite{2022Ahn01} The RIBF accelerator complex at the RIKEN Nishina Canter was used to deliver a 345 MeV/nucleon $^{48}$Ca beam to a 20-mm thick beryllium target. Projectile fragments were separated with the large acceptance two-stage separator BigRIPS and identified by their time of flight, magnetic rigidity, and energy loss measurements. ``We observed nine events for $^{39}$Na and thus established that it is particle-bound.'' A single $^{39}$Na event had been previously reported but it was not deemed sufficient evidence for its discovery.\cite{2019Ahn01}

Yu. Ts. Oganessian et al. reported the discovery of $^{286}$Mc in ``New isotope $^{286}$Mc produced in the $^{243}$Am+$^{48}$Ca reaction''.\cite{2022Oga02} The DC280 cyclotron at the SHE factory complex in Dubna, Russia was utilized to accelerate a $^{48}$Ca to 259.1 MeV which impinged on an enriched $^{243}$Am target. $^{286}$Mc was produced in the 5n evaporation reaction $^{243}$Am($^{48}$Ca,5n)$^{286}$Mc and detected with the gas-filled separator DGFRS-2. Residue and subsequent correlated $\alpha$ decays were measured in a double-sided strip detector and by eight side detectors. ``The new isotope $^{286}$Mc was synthesized, and its half-life of 20$^{+98}_{-9}$ ms and $\alpha$-particle energy of 10.71$\pm$0.02 MeV were determined.''

\section{Discoveries not yet published in refereed journals}

\begin{table}[t]
\tbl{Nuclides only reported in proceedings or internal reports until the end of 2022. The nuclide, first author, reference and year of proceeding or report are listed. \label{reports}}
{\begin{tabular}{@{}llrr@{}} \toprule
\parbox[t]{4.8cm}{\raggedright Nuclide(s) } & \parbox[t]{2.3cm}{\raggedright First Author} & Ref. & Year \\ \colrule

$^{21}$C	&	 S. Leblond 	&	\refcite{2015Leb01,2015Leb02}	&	2015	 \\ 
			&	 N. A. Orr 	&	\refcite{2016Orr01}	&	2016	 \\ 
$^{24}$N, $^{25}$N		&	 Q. Deshayes 	&	\refcite{2018Des01}	&	2018	 \\ 
$^{45}$Si, $^{46}$Si		&	 H. Suzuki 	&	\refcite{2021Suz01}	&	2021	 \\ 
$^{79}$Co, $^{84}$Cu & Y. Shimizu & \refcite{2018Shi02} & 2018 \\
$^{86}$Zn$^a$, $^{88}$Ga, $^{89}$Ga, $^{91}$Ge, $^{93}$As$^a$, 	&	 Y. Shimizu 	&	\refcite{2015Shi01}	&	2015	 \\ 
$^{94}$As, $^{96}$Se, $^{97}$Se, $^{99}$Br, $^{100}$Br & & & \\
$^{98}$Sn	&	  I. Celikovic 	&	\refcite{2013Cel01}	&	2013	 \\ 
$^{155}$Ba, $^{159}$Ce, $^{164}$Nd, $^{173}$Gd, $^{175}$Tb, 	&	 N. Fukuda 	&	\refcite{2015Fuk01}	&	2015	 \\
$^{177}$Dy, $^{179}$Ho, $^{181}$Er, $^{182}$Tm, $^{183}$Tm& & & \\
$^{126}$Nd, $^{136}$Gd, $^{138}$Tb, $^{143}$Ho$^b$, $^{153}$Hf	&	 G. A. Souliotis 	&	\refcite{2000Sou01}	&	2000	 \\
$^{143}$Er, $^{144}$Tm	&	 R. Grzywacz 	&	\refcite{2005Grz01}	&	2005	 \\
	& K. Rykaczewski & \refcite{2005Ryk01} & 2005 \\
	& C. R. Bingham & \refcite{2005Bin01} & 2005 \\
$^{230}$At, $^{232}$Rn	&	 J. Benlliure 	&	\refcite{2010Ben02}	&	2010	 \\
					&	  	&	\refcite{2015Ben01}	&	2015	 \\
$^{252}$Bk, $^{253}$Bk	&	 S. A. Kreek 	&	\refcite{1992Kre01}	&	1992	 \\
$^{262}$No 	&	 R. W. Lougheed 	&	\refcite{1988Lou01},\refcite{1989Lou01}	&	1988/89	 \\
	&	 E. K. Hulet 	&	\refcite{1989Hul01}	&	1989	 \\
$^{261}$Lr, $^{262}$Lr	&	 R. W. Lougheed 	&	\refcite{1987Lou01}	&	1987	 \\
	&	 E. K. Hulet 	&	\refcite{1989Hul01}	&	1989	 \\
	&	 R. A. Henderson 	&	\refcite{1991Hen01}	&	1991	 \\
$^{255}$Db	&	 G. N. Flerov	&	\refcite{1976Fle01}	&	1976	 \\
		& A.-P. Lepp\"anen	& \refcite{2005Lep01} & 2005 \\
\botrule
\vspace*{-0.2cm} & & & \\
$^a$ also published in ref. \refcite{2018Shi02} \\
$^b$ also published in ref. \refcite{2003Sew02} \\
\end{tabular}}
\end{table}

Table \ref{reports} lists the nuclides which so far have only been presented in conference proceedings or internal reports. This includes almost 30 nuclides produced by projectile fragmentation and projectile fission at RIBF which should be published in refereed journals in the near future. Compared to the 2018 update\cite{2019Tho01} two new nuclides ($^{45,46}$Si\cite{2021Suz01}) were added, while eleven nuclides have been published in refereed journals since. One of them ($^{150}$Yb) -- originally reported in a conference proceeding by Souliotis\cite{2000Sou01} -- was the first nuclide ever discovered at TRIUMF in Canada. It was produced by spallation reactions and identified with the TITAN facility.\cite{2021Bec01} In the future this method could also potentially be used to discover the other medium-mass proton-rich isotopes mentioned by Souliotis and listed in the table.

\section{Status at the end of 2022 -- Summary}

\begin{figure}[pt]
\centerline{\psfig{file=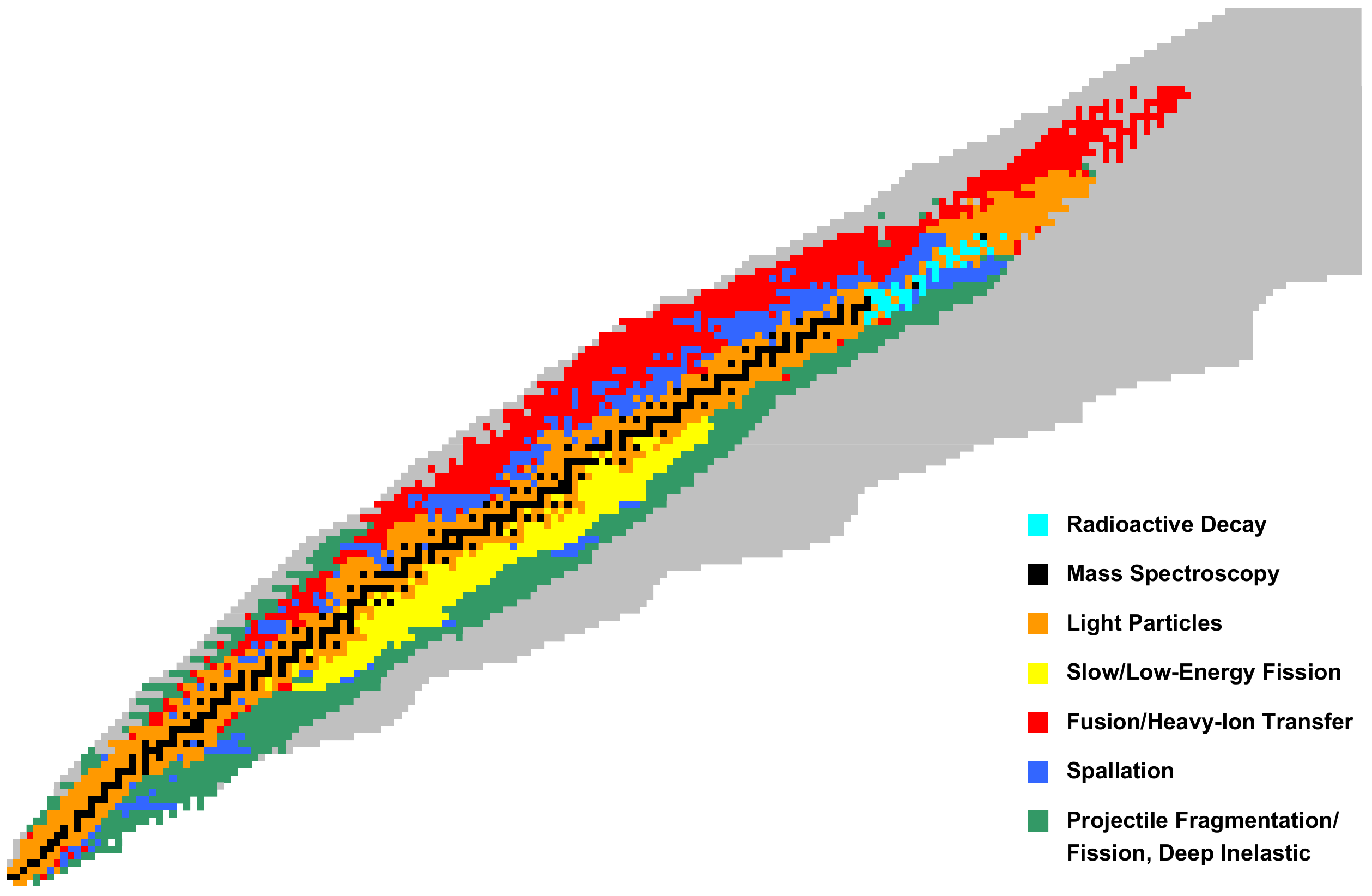,width=12.4cm}}
\caption{Chart of Nuclides. The color code represents the experimental method used to discover the nuclides and is displayed in the figure. The gray area corresponds to undiscovered nuclides which are predicted to exist.\protect\cite{2012Erl01}  \label{f:timeline} }
\vspace*{-0.1cm}
\end{figure}

The 36 new discoveries from 2019 to 2022 increased the total number of observed isotopes to 3338. They were reported by 928 different first authors in 1575 papers and a total of 3969 different coauthors. Further statistics can be found on the discovery project website.\cite{2011Tho03}

Figure \ref{f:timeline} shows the chart of nuclides color-coded by the method of discovery as displayed in the figure legend. An animated version of this chart as a function of year of discovery can also be found at the discovery project website.\cite{2011Tho03} There are still a large number of undiscovered nuclides which are predicted to exist\cite{2012Erl01} as shown by the gray area in the figure. Clearly the largest discovery potential is located in the neutron-rich area of the chart where essentially all recent discoveries were made using projectile fragmentation or projectile fission reactions. Thus, the newly constructed Facility for Rare Isotope Beams at Michigan State University\cite{2022Wei01,2022Wei02} will discover many more new nuclides in the near future as it reaches full beam intensity.\cite{2022McG01}



\bibliographystyle{ws-ijmpe}
\bibliography{isotope-references-etal}

\end{document}